%% file: main.tex
\documentclass[runningheads]{llncs}
\usepackage{graphicx}
\usepackage{amsmath}
\usepackage{amsfonts,amssymb}
\usepackage{multirow}
\usepackage{booktabs}
\pdfoutput=1
\makeatletter 
\newcommand\tabcaption{\def\@captype{table}\caption} 
\makeatother

\begin{document}

 \title{
     Stabilize, Decompose, and Denoise:  Self-Supervised Fluoroscopy Denoising
}
\author{
    Ruizhou Liu\inst{1}\orcidID{0000-0002-2940-9630}
    \and Qiang Ma\inst{1}\orcidID{0000-0003-0791-1731}
    \and Zhiwei Cheng\inst{1}\orcidID{0000-0002-1980-811X}
    \and Yuanyuan Lyu\inst{1}\orcidID{0000-0001-9049-7812}
    \and Jianji Wang\inst{2}\orcidID{0000-0001-5373-6247} 
    \and S. Kevin Zhou\inst{3}\orcidID{0000-0002-6881-4444}
}
\institute{
Z2Sky Technologies Inc. \and Affiliated Hospital of Guizhou Medical University \and University of Science and Technology of China\\
\email{s.kevin.zhou@gmail.com}
}

\authorrunning{R. Liu et al.}

\maketitle              

\begin{abstract}
Fluoroscopy is an imaging technique that uses X-ray to obtain a real-time 2D video of the interior of a 3D object, helping surgeons to observe pathological structures and tissue functions especially during intervention. However, it suffers from heavy noise that mainly arises from the clinical use of a low dose X-ray, thereby necessitating the technology of fluoroscopy denoising. Such denoising is challenged by the relative motion between the object being imaged and the X-ray imaging system. We tackle this challenge by proposing a self-supervised, three-stage framework that exploits the domain knowledge of fluoroscopy imaging. (i) Stabilize: we first construct a dynamic panorama based on optical flow calculation to stabilize the non-stationary background induced by the motion of the X-ray detector. (ii) Decompose: we then propose a novel mask-based Robust Principle Component Analysis (RPCA) decomposition method to separate a video with detector motion into a low-rank background and a sparse foreground. Such a decomposition accommodates the reading habit of experts. (iii) Denoise: we finally denoise the background and foreground separately by a self-supervised learning strategy and fuse the denoised parts into the final output via a bilateral, spatiotemporal filter. To assess the effectiveness of our work, we curate a dedicated fluoroscopy dataset of 27 videos (1,568 frames) and corresponding ground truth. Our experiments demonstrate that it achieves significant improvements in terms of denoising and enhancement effects when compared with standard approaches. Finally, expert rating confirms this efficacy. 

\end{abstract}

\section{Introduction}
Fluoroscopy is a medical imaging technique that uses X-ray to monitor the interior structure of human body in real-time. It helps surgeons to observe pathological structures and tissue functions especially during intervention, without destroying the external epidermis. While it is desirable to reduce a radiation dose for less harm in clinical practice, the use of low dose results in heavy noise in raw fluoroscopy, thereby necessitating the technology of fluoroscopy denoising. But, fluoroscopy denoising is challenging due to the relative motion of the object being imaging with respect to the imaging system as well as a lack of ground truth clean data.

Video denoising, one of the most fundamental tasks in computer vision, is a procedure that eliminates measurement noise from corrupted video data and recovers the original clean information. In mathematical terms, a corrupted video $y$ can be represented as $y = x + n$, where $x$ is clean video data and $n$ is measurement noise. 
Conventionally, not only edge-preserving adaptive filters\cite{T.Cerciello2011}, non-local means\cite{2011nlm,7098781,4271520,vbm4d} denoising methods, but Robust Principle Components Analysis (RPCA)~\cite{candes2011robust,feng2013online,10.1007/978-3-030-87234-2_56,he2012incremental,moore2019panoramic,rodriguez2016incremental}, a prominent method on foreground-background problem, is often used on video denoising task, which uses a low-rank subspace
model to estimate the background and a spatially sparse model to estimate the foreground. Some models~\cite{Guyon2012ForegroundDV,inproceedingsForeground_Detection} employing Total Variation (TV) on RPCA to separate foreground and background, like TVRPCA~\cite{TVRPCA}. While Inc-PCP\cite{rodriguez2016incremental} can iteratively align the estimated background component. However, for non-static background video, a RPCA-based~\cite{ErfanianEbadi2016ApproximatedRP} model was proposed, but the approach considers only the common view of the video. In addition, when processing such video data with the Inc-PCP method, the decomposed background information is lost due to the moving background.

Recently, deep learning based denoising approaches become prominent. Fully supervised methods usually train a neural network with \textit{corrupted/clean data pairs}~\cite{claus2019videnn,wang2020first,zhang2018ffdnet}. However, they require collecting paired data and also have the likelihood of learning an identity mapping instead of statistic feature of noise. Self-supervised models aim to avoid this issue by either inferring real data by the local perceived corrupted data, or predicting clean images based on characteristics of noise~\cite{batson2019noise2self,ehret2019model,krull2019noise2void,lehtinen2018noise2noise,quan2020self2self}.
Meanwhile, most approaches are designed for natural images, while denoising algorithms for medical imaging data are less common.
 
In this paper, we propose a novel self-supervised, three-stage fluoroscopy denoising framework 
for processing a fluoroscopy video with a non-static background. In the first stage, \textbf{Stabilize}. In order to preserve temporal consistency and stabilize the given fluoroscopy imaging data, which might still undergo a global motion, a dynamic panorama is constructed for frame registration. In the second stage, \textbf{Decompose}. We proposed a simple but novel mask-based RPCA decomposition method to separate the given imaging video with detector motion into a low-rank background and a sparse foreground.  In the third stage, \textbf{Denoise.} The background and foreground will be denoised, respectively, with a self-supervised denoising strategy, and the foreground is denoised again with spatiotemporal bilateral filtering. Finally, these parts are fused together to obtain final result. We improve the denoising performance of our framework by 5$\sim$6 dB on a dedicated fluoroscopy imaging video dataset comparing with other approaches. Expert evaluations confirm the efficacy of our framework on clinical images too.

\begin{figure*}[t]
\centering \vspace{-0.2cm}
    \includegraphics[width=\textwidth]{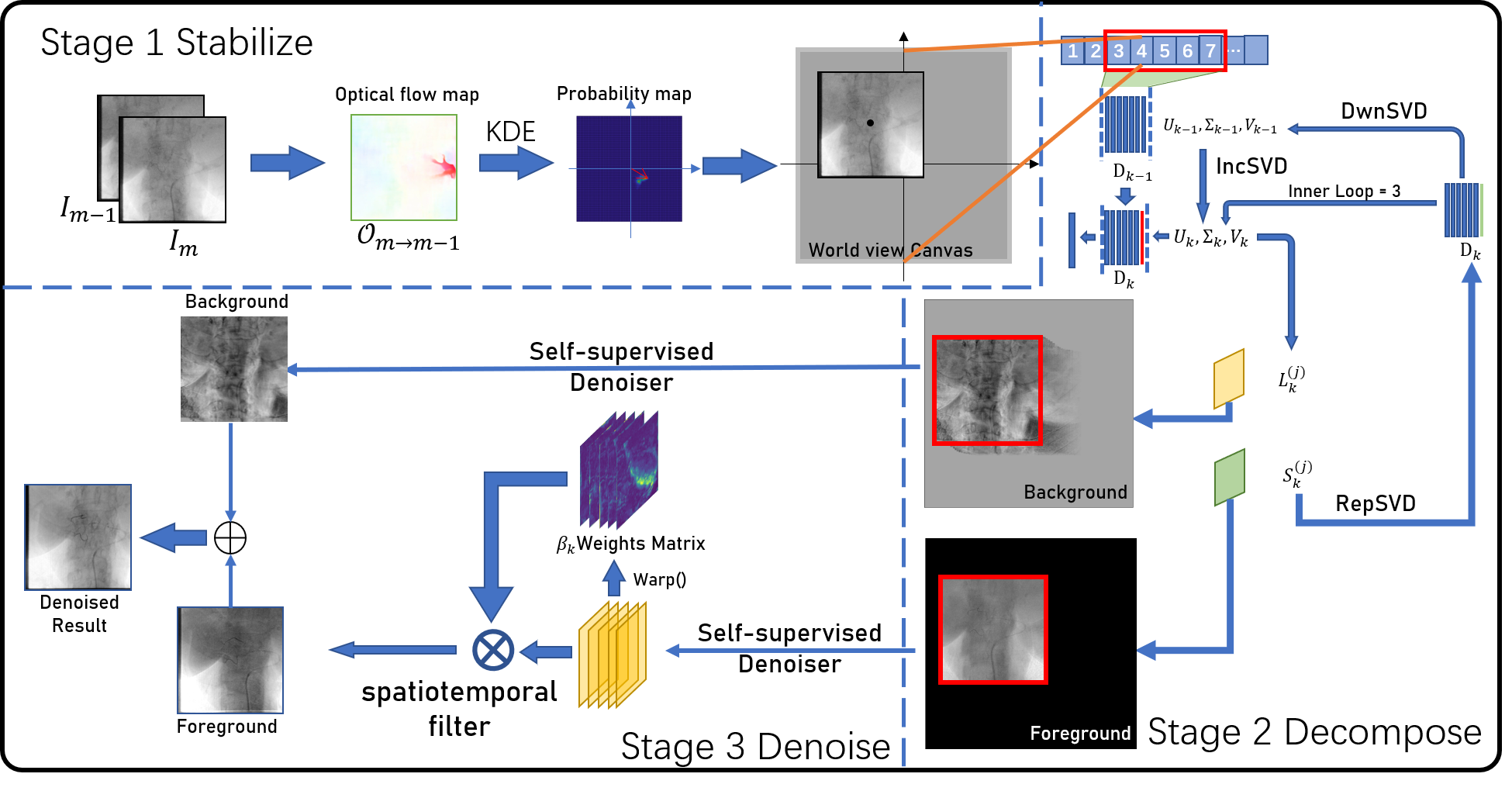}
    \caption{Overview of the Stabilize, Decompose, and Denoise framework for fluoroscopy denoising.} \label{fig1}
\vspace{-0.4cm}\end{figure*}

\vspace{-0.2in}\section{Method}\vspace{-0.2cm}

We design a self-supervised framework for reducing the noise in fluoroscopy, tackling problem of background shift in RPCA decomposition, and supporting fast data processing. As in Fig.~\ref{fig1}, our framework consists of three stages --- Stabilize, Decompose and Denoise, whose details are elaborated below.

\vspace{-0.2cm}\subsection{Stage 1 - Stabilize}\vspace{-0.2cm}
In a fluoroscopy video, the motion of X-ray detector results in a shift in the background, which makes temporal denoising challenging. Inspired by \cite{moore2019panoramic}, we construct a panorama on a world view canvas for each input video frame to preserve the temporal consistency of input video by compensating the translation between frames.

For a given video stream ${Y:=\{Y_{0},Y_{1}, ...,Y_{T-1}\} \in \mathbb{R}^{M\times N\times T}}$, 
we first construct a world view canvas. We use the first frame ${Y_{0}}$ as the reference frame of the entire video and place it at the center of the world view canvas. 
When any given frame ${Y_{m}}$ is fed into this stage, we calculate an optical flow field ${\mathcal{O}_{m\rightarrow m-1} \in \mathbb{R}^{M\times N\times 2}}$ between ${Y_{m-1}}$ and ${Y_{m}}$ with optical flow estimator ${\mathcal{G}_{flow}}$,
which is implemented by PWC-Net~\cite{sun2018pwc}.

Though it is likely that the video foreground possesses more pixels than background, the motion patterns of foreground pixels are random, while those of background pixels are consistent. Thus, the distribution of the optical flow field ${\mathcal{O}_{m\rightarrow m-1}}$ is estimated using the kernel density estimator (KDE), then the value of ${[u_{max}, v_{max}]}$ with a maximum probability density value is selected as the background offset between the current and previous frames. Finally, we can obtain a panoramic current frame $Y_{m}^{pano}$ by compensating the translation transformation. 
Each frame is placed on their appropriate position and temporal consistency of video is preserved.

\vspace{-0.2cm}\subsection{Stage 2 - Decompose}\vspace{-0.2cm}

For a given frame $Y_i$, it can be decomposed into sparse foreground $S_i$ and low-rank background $L_i$ according to RPCA assumption. Then, given a mask ${{\mathcal{M}_i} \in \mathbb{R}^{MN \times 1}}$ of $S_i$, in which ones elements in the ${\mathcal{M}_i}$ indicats non-zero elements of ${{{S}_i}}$, so we have ${\|{\mathcal{M}_i}\|_{F} \ll \|1-{\mathcal{M}_i}\|_{F}}$, and the $\|\cdot\|_{F}$ is Frobenius norm. Since the component $L_i$ complies with the assumption of RPCA, it does not change significantly along the temporal axis, which means that most of the noise energy has been accumulated on the foreground. Therefore, 
if we denoise the background ${L}_i$ to obtain ${\hat{L}_i}$ and denoise the foreground ${S_i}$ to obtain $\hat{S}_i$, then noise energy of $\hat{Y}_i={\hat{L}_i}+\hat{S}_i$ must be no more than that of ${\hat{{X}}_i}$, which is the result obtained by directly denoising the corrupted image ${{Y}_i}$. Assuming that ${X}_i$ is a clean version of ${{Y}_i}$, we can prove (in supplementary materials) that
\begin{equation}\label{eq:L}
    {\|
    (\boldsymbol{1}-{\mathcal{M}_i})
    \odot
    ({X_i}-\hat{{L}}_i)
    \|_{F}^{2}
    \leq
    \|
    (\boldsymbol{1}-{\mathcal{M}_i})
    \odot
    ({X}_i-\hat{{X}}_i)
    \|_{F}^{2}},
\end{equation}
where the $\odot$ is element-wise multiplication. Denoting the errors in the denoised results ${\hat{Y}_i}$ and ${\hat{X}_i}$ by $\epsilon({X}_i, {\hat{Y}_i})$ and $\epsilon({X}_i, {\hat{X}}_i)$, respectively, using (\ref{eq:L}), we can further prove that $\epsilon({X}_i, {\hat{Y}_i}) \leq \epsilon({X}_i, {\hat{X}_i})$ (refer to supplementary materials for more details), which means that \textbf{video decomposition results in better denoising performance}. 

However, since two consecutive frames after translation compensation do not perfectly overlap with each other in the canvas, we need to deal with this issue. We use ${\mathcal{P}_{M}}$ and ${\mathcal{P}_{\bar M}}$ to indicate non-overlapped area and overlapped area between $Y_{i-1}$ and $Y_{i}$, respectively. Generally speaking, the non-overlapped area introduces new information, which makes ${\mathcal{P}_{M}({S}_i)=0}$ and ${\mathcal{P}_{M}({L}_i)=\mathcal{P}_{M}({Y}_i)}$. Therefore, we have
\vspace{-0.1cm}
\begin{equation} \label{eq:L2}  \vspace{-0.1cm}
   \begin{cases}
        \mathcal{P}_{\bar M}(\|
        (\boldsymbol{1}-{\mathcal{M}_i})
        \odot
        ({X}_i-\hat{L}_i)
        \|_{F}^{2})
        \leq
        \mathcal{P}_{\bar M}(\|
        (\boldsymbol{1}-{\mathcal{M}_i})
        \odot
        ({X}_i-\hat{X}_i)
        \|_{F}^{2}), \\
        \mathcal{P}_{M}(\|
        {X}_i-\hat{L}_i
        \|_{F}^{2})
        =
        \mathcal{P}_{M}(\|
        {X}_i-\hat{X}_i
        \|_{F}^{2}).
    \end{cases} 
\end{equation}
With the help of (\ref{eq:L2}), the same statement $\epsilon({X}_i, {\hat{Y}_i}) \leq \epsilon({X}_i, {\hat{X}_i})$ still holds (again refer to supplementary materials).

Because the conventional RPCA decomposition methods cannot tackle aforementioned problem, such as Inc-PCP\cite{rodriguez2016incremental}, we proposed mask-based RPCA decomposition method as decomposition module in the framework. It inherits four operations from Inc-PCP, PartialSVD, IncSVD, RepSVD and DwnSVD. Through these operations, the $U$ matrix reserving previous background information and two weight matrices $\Sigma$ and $V$ in the algorithm are maintained and updated. However, because the positions of two consecutive panoramic frames on world view are not the same, we need to fill the non-overlapped areas on $Y_i$ and $U$ before decomposition using the below equations:.
\begin{equation}\label{fillY}
\begin{cases}
    \mathcal{P}_{M_{Y}}(Y_i) = \mathcal{P}_{M_{Y}}(U)\text{diag}(\Sigma)(\mathcal{P}_{\bar M}(U)\text{diag}(\Sigma))^{+}\mathcal{P}_{\bar M}({Y_i});
    \\
    \mathcal{P}_{M_{U}}(U) =  \mathcal{P}_{M_{U}}(Y_i)\mathcal{P}_{\bar M}(Y_i)^{+}(\mathcal{P}_{\bar M}(U)\text{diag}(\Sigma))\text{diag}(\Sigma)^{+},
    \end{cases}
\end{equation}
where $\mathcal{P}_{M_{U}}$ indicates unknown area on $U$ relatively to known area on ${Y_i}$, and $\mathcal{P}_{M_{Y}}$ refers as unknown area on ${Y_i}$ corresponding to known area on $U$. The $(\cdot)^{+}$ is pseudo inverse matrix. The first equation of (\ref{fillY}) is used to complete non-overlapped area on ${Y_i}$, and the second equation of (\ref{fillY}) is used to fill non-overlapped blank area on $U$. The aforementioned Inc-PCP operations are then used to decompose ${Y_i}$ into the foreground $S_i$ and background $L_i$ and update the $U$ matrix, $\Sigma$ matrix and $V$ matrix.

\vspace{-0.2cm}\subsection{Stage 3 - Denoise}\vspace{-0.2cm}
At this stage, a single-frame self-supervised denoising network $\mathcal{F}_{\theta}(\cdot)$ is deployed to denoise the background ${L_i}$ and foreground ${S_i}$ decomposed from a corrupted frame ${Y_i}$, and the denoised results are denoted by ${\hat{L}_i}$ and ${\hat{S}_i}$, respectively. In this work, the Self2Self~\cite{quan2020self2self} denoising network is used, and it is worth noting that the Denoise stage is a general stage for self-supervised denoising, which means the denoising model can be substituted with other self-supervised denoising approaches. In the training phase of the Self2Self network, we first generate a lot of training samples through Bernoulli sampling  $\boldsymbol{x}^i\sim\mathcal{X}$, donated as $\{{\hat{\boldsymbol{x}}^i}_{m}\}_{m=1}^{M}$ ,where ${\hat{\boldsymbol{x}}^i}_m:=\boldsymbol{b}_m \, \odot \, \boldsymbol{x}^i$. Next, let ${\overline{\boldsymbol{x}}^i_{m}:=(\boldsymbol{1}-\boldsymbol{b}_{m}) \odot \boldsymbol{x}^i}$ and $\boldsymbol{b}_m$ is a down sampling mask. The network is then trained by minimizing the following loss function 
\vspace{-0.3cm}
\begin{equation} \vspace{-0.1cm}
    \mathcal{L}(\theta)=\mathbb{E}_{\boldsymbol{x\sim \mathcal{X}}}\left[\sum\limits_{m=1}^{M}\|
    \mathcal{F}_{\theta}(\boldsymbol{\hat{x}}_m)-\boldsymbol{\overline{x}}_m
    \|_{\boldsymbol{b_{m}}}^{2}\right].
\end{equation}

But in the testing stage, for reducing prediction time, a U-Net~\cite{ronneberger2015u} as a student module $\mathcal{D}_\omega(\cdot)$ is added into the framework, learning the features of Self2Self network in a fully-supervised manner. The loss function $\mathcal{L}_{student}(\omega)$ is given as
\begin{equation}\vspace{-0.1cm}
    \mathcal{L}_{student}(\omega)=\mathbb{E}_{\boldsymbol{x\sim \mathcal{X}}}\left[\sum\limits_{m=1}^{M}\|
    \mathcal{D}_\omega(\boldsymbol{x})-\mathcal{F}_{\theta}(\boldsymbol{x})
    \|_{F}^{2}\right].
\end{equation}

Finally, an optical flow based spatiotemporal bilateral filter on foregrounds is adopted for multi-frame denoising. Given $2K+1$ consecutive foregrounds ${\hat{S}}_{t-K},...,{\hat{S}}_{t},...,{\hat{S}}_{t+K}$, for each $k \in \{-K,...,K\}$, the optical flow $\mathcal{O}_{t+k \rightarrow t} \in \mathbb{R}^{M\times N\times 2}$ is calculated. 
Then we use a bilateral filter to average the warped $2K+1$ frames for denoising $\overline{{S}}_t$,
\begin{align}
    &\overline{{S}}_t=\sum_{k=-K}^{K} \beta_{k} \cdot 
    \mathcal{W}({\hat{S}}_{t+k}, \mathcal{O}_{t+k \rightarrow t}), \\
    & 
    \beta_{k} \propto exp\{-(\mathcal{W}({\hat{S}}_{t+k}, \mathcal{O}_{t+k \rightarrow t})-{\hat{S}}_{t})/\rho\}, \sum_k \beta_{k} =1, 
\end{align}
where $\mathcal{W}(\cdot)$ is a warping function and $\rho$ is a parameter controlling the smoothness, with a larger $\rho$ value producing a smoother denoising result.

\section{Experiment}
\vspace{-0.2cm}\subsection{Setup Details}\vspace{-0.1cm}
\textbf{Fluoroscopy dataset.} We collect 27 clean fluoroscopy videos (1,568 frames in total) with a high X-ray dose as ground truth, including 20 static-background videos and 7 non-static background videos. For evaluation, we add Gaussian noise with different variances of 0.001, 0.003, and 0.005 into clean images to simulate corrupted data with different noise levels.

\textbf{Clinical dataset.} We collect 60 groups of real corrupted samples by sampling low dose X-ray videos. Each group consists of 5 images, one of which is the original noisy image, and the remaining four are denoised results obtained by feeding the original noisy image into our method, Noise2Self, Noise2Void and Self2Self, respectively. In addition, for each group, the five images are permuted randomly.

\textbf{Implementation and training stage.} We adopt the Self2Self as denoiser, which is trained on X-ray Coronary Angiograms dataset (XCA) of 22 videos and set the probability of Bernoulli sampling as 0.3. When we train student network, the Adam optimizer is used and the learning rate is set as $10^{-4}$, and it is trained with 100 epochs with a batch size of 8. In the Stabilize stage, the size of world view canvas is $2048\times2048$ and frame size $M\times N$ is $1024\times1024$. The Inc-PCP algorithm is implemented on GPU except the part of filling blank area, with the rank ${r=1}$, and the windows size is 30. In spatiotemporal bilateral filter, we set $\rho=0.02$. All experiments are implemented on an NVIDIA GeForce RTX 2080Ti GPU and using PyTorch.

\textbf{Metrics.} We use peak signal-to-noise ratio (PSNR) and structural similarity index (SSIM) to evaluate denoising performance by comparing the denoised images and its corresponding clean images. To evaluate the extend of blurriness for a given image, image entropy (IE) is used for quantification.

\textbf{Expert rating.} A proficient radiologist with about 20 years of reading experience is invited to rate the denoised image quality for our Clinical dataset. The criteria include perceived noisiness and completeness of micro-structures like small vessels. The radiologist is asked to rate all the images from each group presented in a random order, from 1 (bad) to 5 (good).

\input{tables/denoise-gene}

\subsection{Results and Discussion}\vspace{-0.2cm}
\textbf{Restoration}. We select some state-of-the-art self-supervised deep models, including Noise2Void\cite{krull2019noise2void}, Noise2Self\cite{batson2019noise2self}, Self2Self\cite{quan2020self2self} for comparison. The comparison performances are demonstrated in Table \ref{tab:denoise-gene}. It is evident that the PSNR and SSIM scores of our framework (Ours+S2S, the last column) among these self-supervised denoising methods are the best, contributing an improvement of about 5$\sim$6 dB compared with other self-supervised methods. Especially the comparison between our results and the results of Self2Self (the second to last column) clearly demonstrates the effectiveness of the Stabilize and Decompose stages. Similarly, in terms of SSIM, ours framework records the best performance. It is noted that even when the noise level increases, our framework is robust too. Fig. \ref{fig:show-results} visualizes the processed results of these methods.

\input{tables/denoise-selfsup}
\input{tables/show_result}

\begin{figure*}[!htb]\vspace{-0.2cm}
\centering
    \includegraphics[width=0.85\textwidth]{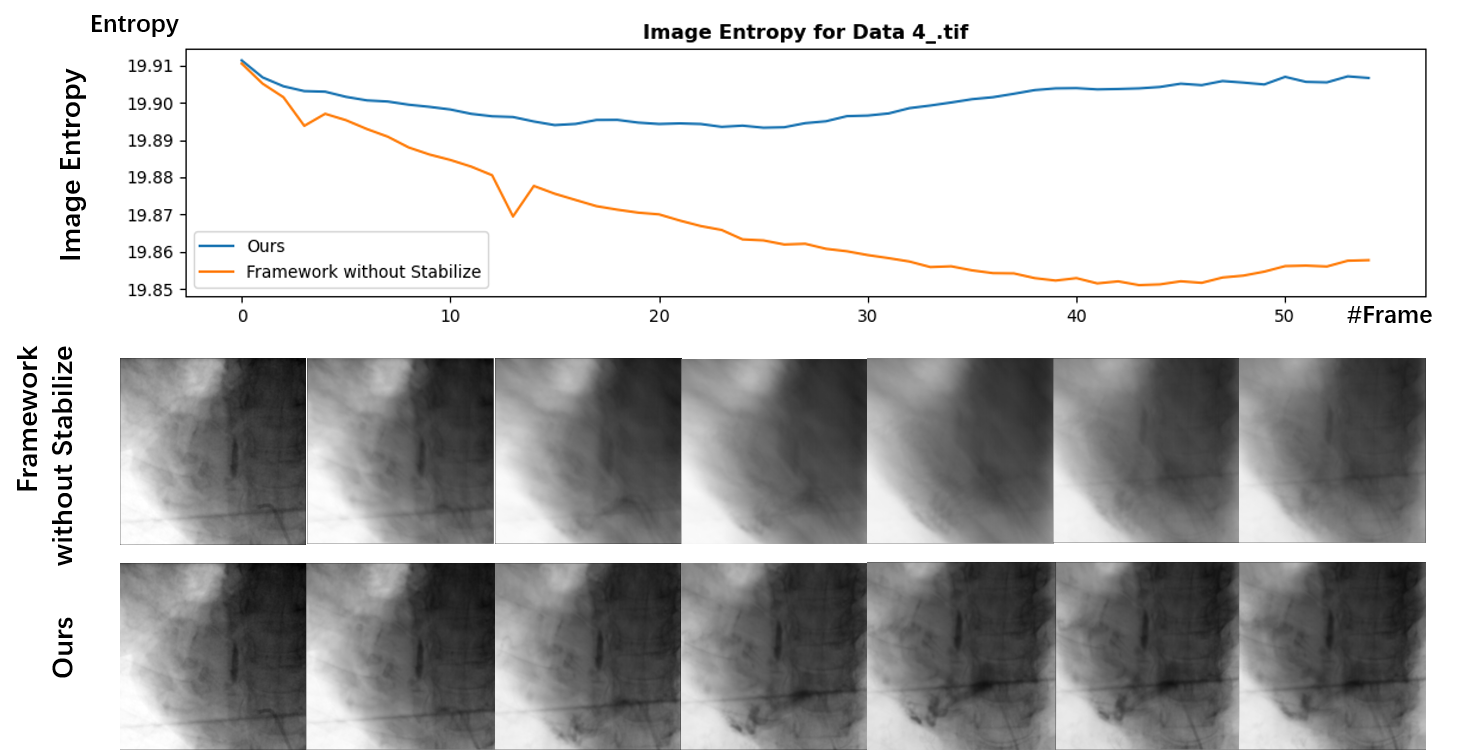}
    \caption{Top: the image entropy curves of backgrounds generated by our framework with and without the Stabilize stage, respectively. Bottom: the visualization of backgrounds for the two results.} 
    \label{fig:background_compare}\vspace{-0.2cm}
\end{figure*}

\textbf{Ablation study}. First of all, the detector motion in fluoroscope yields background shifting, which leads to serious blurriness in the background obtained in video decomposition stage. To quantify this artifact, we calculate the image entropy (IE) for the background to verify the effectiveness of the Stabilize stage for preserving temporal consistency of video. The higher IE is, the better. We plot the IE curves of one fluroscopy video with and without the Stabilize stage as shown in Fig. \ref{fig:background_compare}. It is clear that the existence of the Stabilize stage preserves more information on the background. Table \ref{tab:denoise-selfsup} indicates that preserving more background information can improve the denoising performance. In addition, in order to verify the generalization of our framework, we replace the denoiser in the Denoise stage with other self-supervised models, like Noise2Void, Noise2Self, the result is shown in Table \ref{tab:denoise-gene}. It is clear that our framework boosts a significant performance for all denoising network. 

\begin{figure}[h] 
    \centering
    \begin{minipage}[b]{0.5\textwidth}
        \input{tables/clinical_show}
    \end{minipage}
    \begin{minipage}[b]{0.42\textwidth}
        \input{tables/clinical-table}
    \end{minipage}
\end{figure}

\textbf{Clinical study}. Table \ref{tab:ranking} summarizes the average ratings and P-values for comparison between our model and other competing methods. The performance of our framework is significantly better than Noise2Self\cite{batson2019noise2self}, Noise2Void\cite{krull2019noise2void}, Self-2Self\cite{quan2020self2self} on our clinical dataset. Fig. \ref{fig: clinical} shows one group of denoising results of real corrupted data. The top-left, top-right, bottom-left, and bottom-right are denoised by Noise2Void\cite{krull2019noise2void}, Noise2Self\cite{batson2019noise2self}, Self2Self\cite{quan2020self2self} and our framework, respectively. It is obvious that the denoised image generated by our framework possesses less noise and better preserves the micro-structures in the image.

\section{Conclusion}

We propose a three-stages self-supervised denoising framework, consisting of the Stabilize, Decompose, and Denoise stages. In the Stabilize stage, we firstly estimate the optical flow map and then the background offset for each frame to build a panorama. Then in the second Decompose stage, a mask-based RPCA decomposition method is proposed for separating foreground and background. Finally, we invoke a self-supervised denoising method to denoise foreground and background, respectively, and fuse them together with a bilateral temporal-spatial filter as final denoised result. 
In experiments, visual comparisons and qualitative evaluations demonstrate that our framework yields better image quality than competing methods and exhibits a great potential of boosting self-supervised learning denoising method on Fluoroscopy dataset and Clinical dataset. In the future, we plan to employ deep learning model to estimate affine parameters so that our framework can work on more complex circumstance.


\newpage
\bibliographystyle{splncs04}
\bibliography{main}

\end{document}

%% file: tables/denoise-gene.tex
\begin{table}[t]
\small
\centering \vspace{-0.2cm}
\caption{Denoising performances of various self-supervised denoising methods. In each cell, the PSNR and SSIM values are presented. The italics shows our framework adopted other denoiser in Denoise stage. The \textbf{bold} shows the best scores.}
\resizebox{\textwidth}{!}
{
    \begin{tabular}{|c|c|c|c|c|c|c|}
    \hline
   {
    \multirow{2}{*}{Gaussian}
    } & \multicolumn{6}{c|}{Method}                                                                                                                        \\ \cline{2-7}    
    & Noise2Void     & \textit{Ours w/ N2V}          & Noise2Self     & \textit{Ours w/ N2S}                  & Self2Self     & \textit{Ours w/ S2S}          \\ \hline
    0.001    & 19.27/.919 & { {25.13/\textbf{.960}}}  & 30.69/.935 & {{ {34.78/.948}}} & 34.39/.941 & {{ \textbf{36.47}/.941}}  \\ 
                                 0.003    & 15.74/.856  & { {21.98/\textbf{.920}}}  & 23.99/.854 & {{ {29.80/.895}}} & 29.96/{.870} & {{ \textbf{34.18}}/.864}  \\ 
                                 0.005    & 14.25/.820  & {{19.29/.890}}  & 21.01/.801 & {{ {26.86/.854}}} & 27.72/.819 & {{ \textbf{32.78/.938}}}  \\ \hline 
    \end{tabular}
}

\label{tab:denoise-gene}

\end{table}
\vspace{-0.5cm}

%% file: tables/denoise-selfsup.tex
\begin{table}[t]
\small
\centering \vspace{-0.2cm}
\caption{Denoising performances of our framework with and without the Stabilize stage. In each cell, the PSNR and SSIM values are presented.}
\resizebox{0.75\textwidth}{!}{
\begin{tabular}{|c|c|c|c|c|}
\hline
\multirow{2}{*}{Method}     & \multicolumn{3}{c|}{Gaussian}                 & \multirow{2}{*}{FPS} \\ \cline{2-4}
                                                   & 0.001         & 0.003         & 0.005         &           \\ \hline
Full three-stage  & \textbf{36.466/.941} & \textbf{34.183/.864} & \textbf{32.783/.938} & 0.72                 \\ 
w/o Stabilize & 36.223/.939 & 33.229/.852 & 30.239/.933 & 4.20                  \\ \hline
\end{tabular}
}
\label{tab:denoise-selfsup}
\vspace{-0.4cm}
\end{table}


%% file: tables/show_result.tex

\begin{figure}[!htb]
\small
\centering 

\resizebox{\textwidth}{!}{
    \begin{tabular}{@{}ccccccccl@{}}
    \tiny Input & \tiny GT & \tiny N2V & \tiny N2S & \tiny S2S & \tiny Ours\\
    \includegraphics[width=0.083\textwidth]{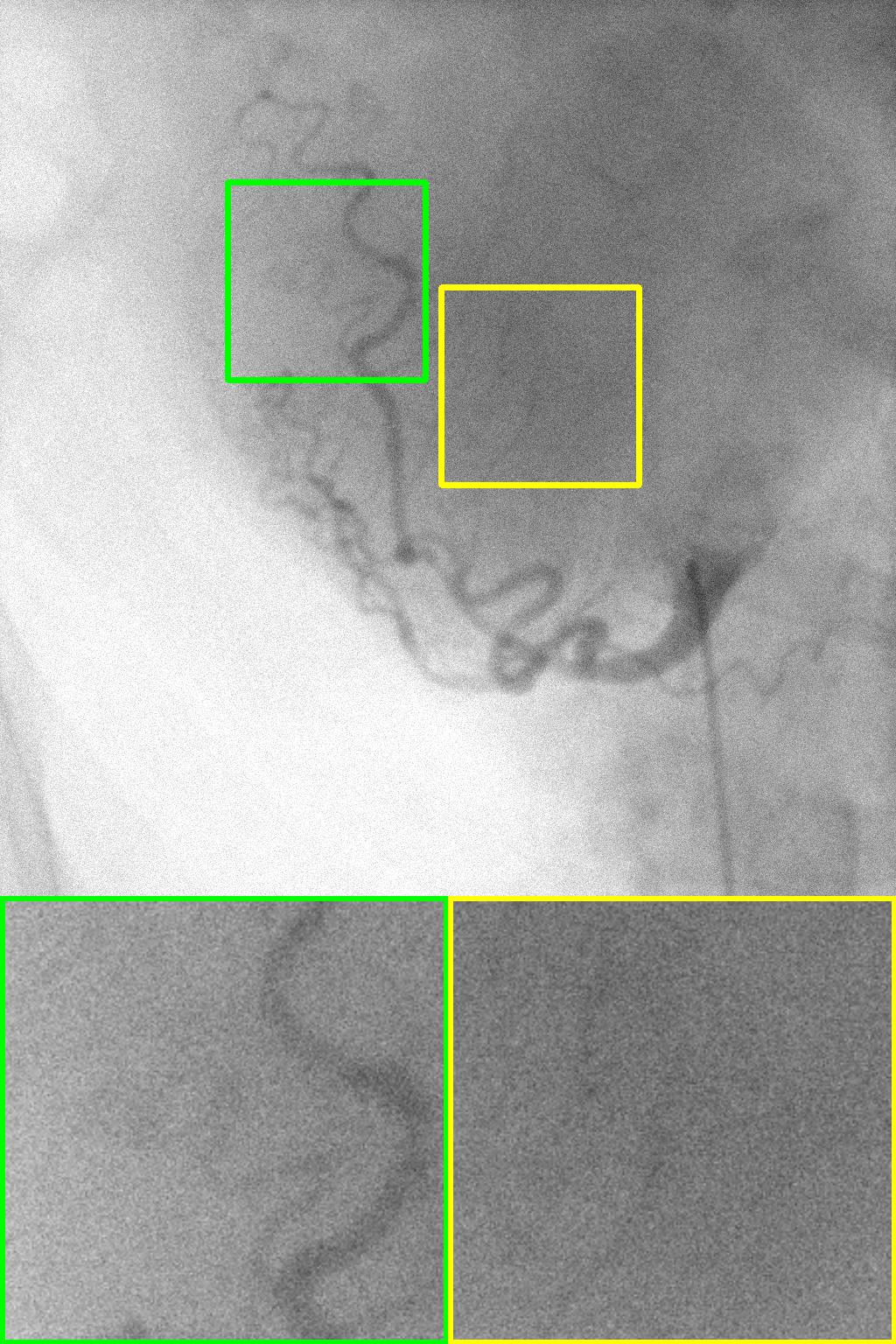} &
    \includegraphics[width=0.083\textwidth]{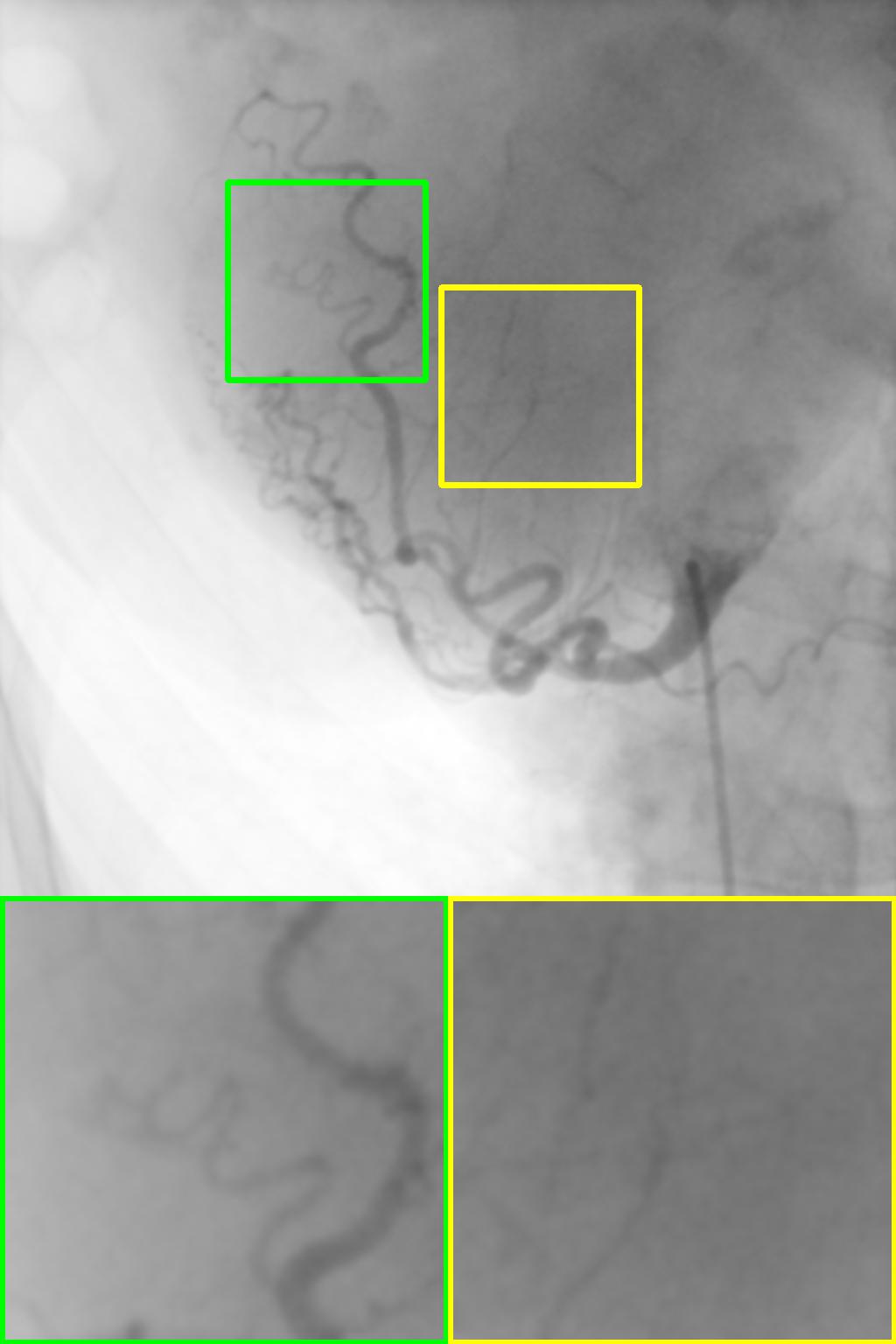} &
    \includegraphics[width=0.083\textwidth]{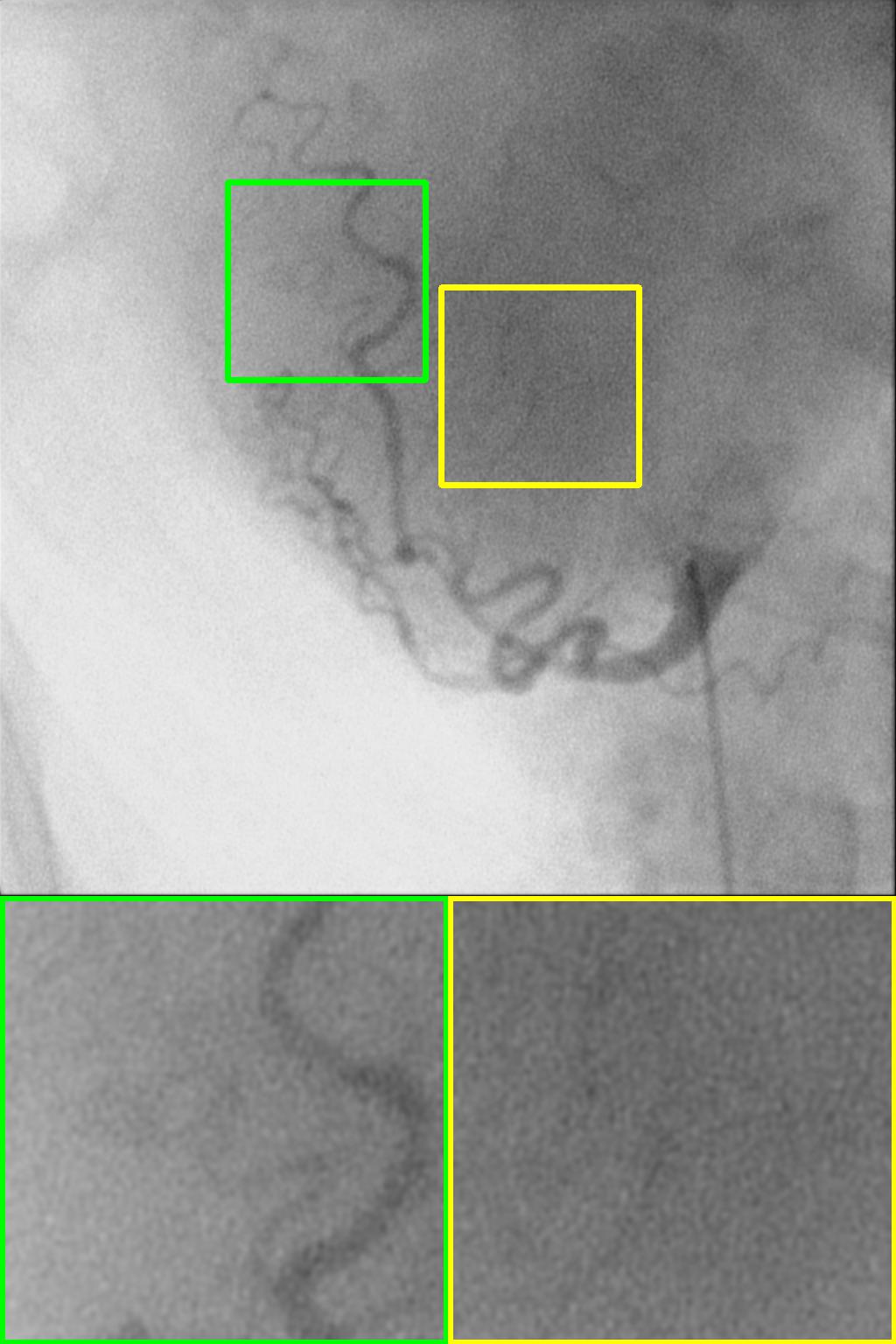} &
    \includegraphics[width=0.083\textwidth]{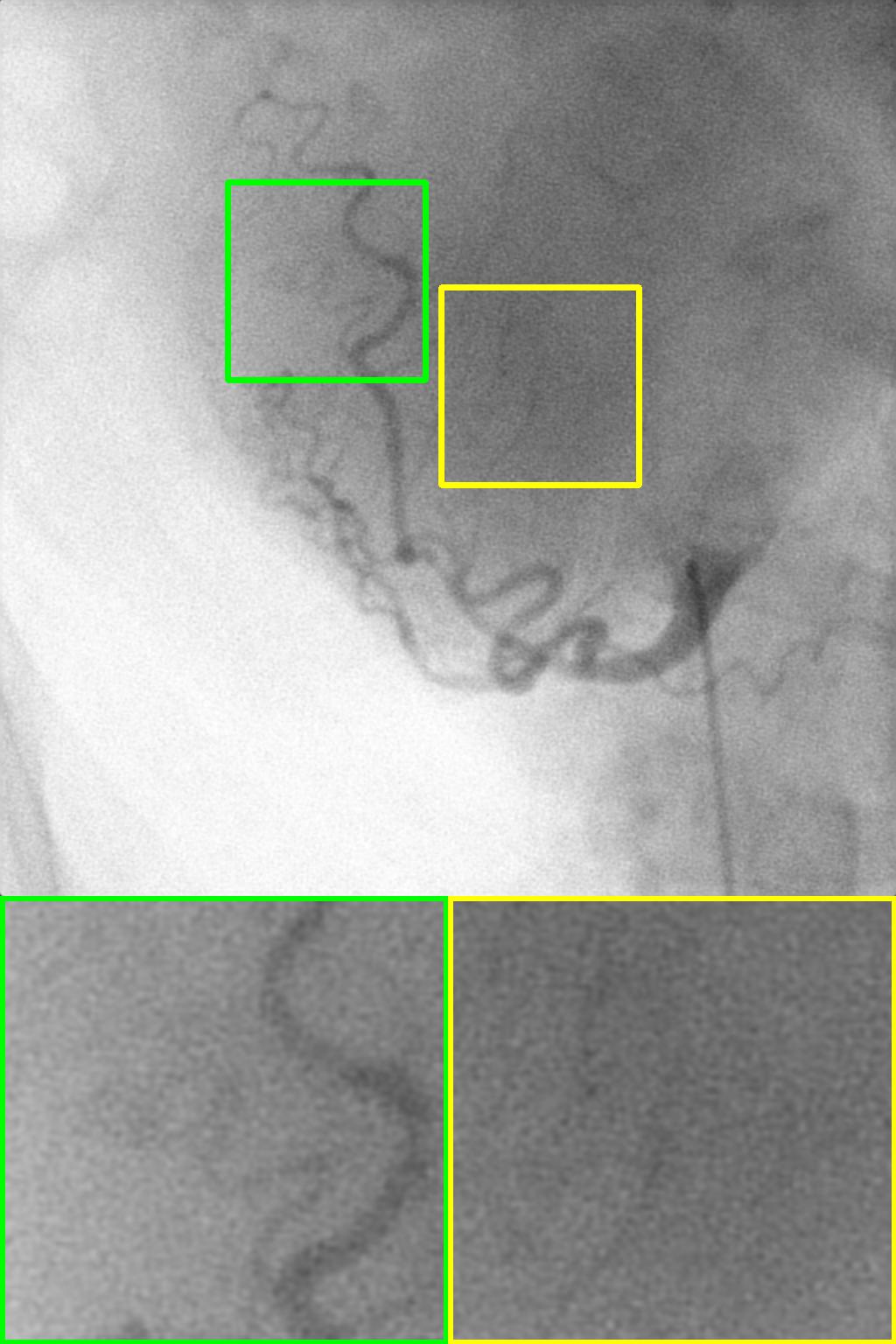}&
    \includegraphics[width=0.083\textwidth]{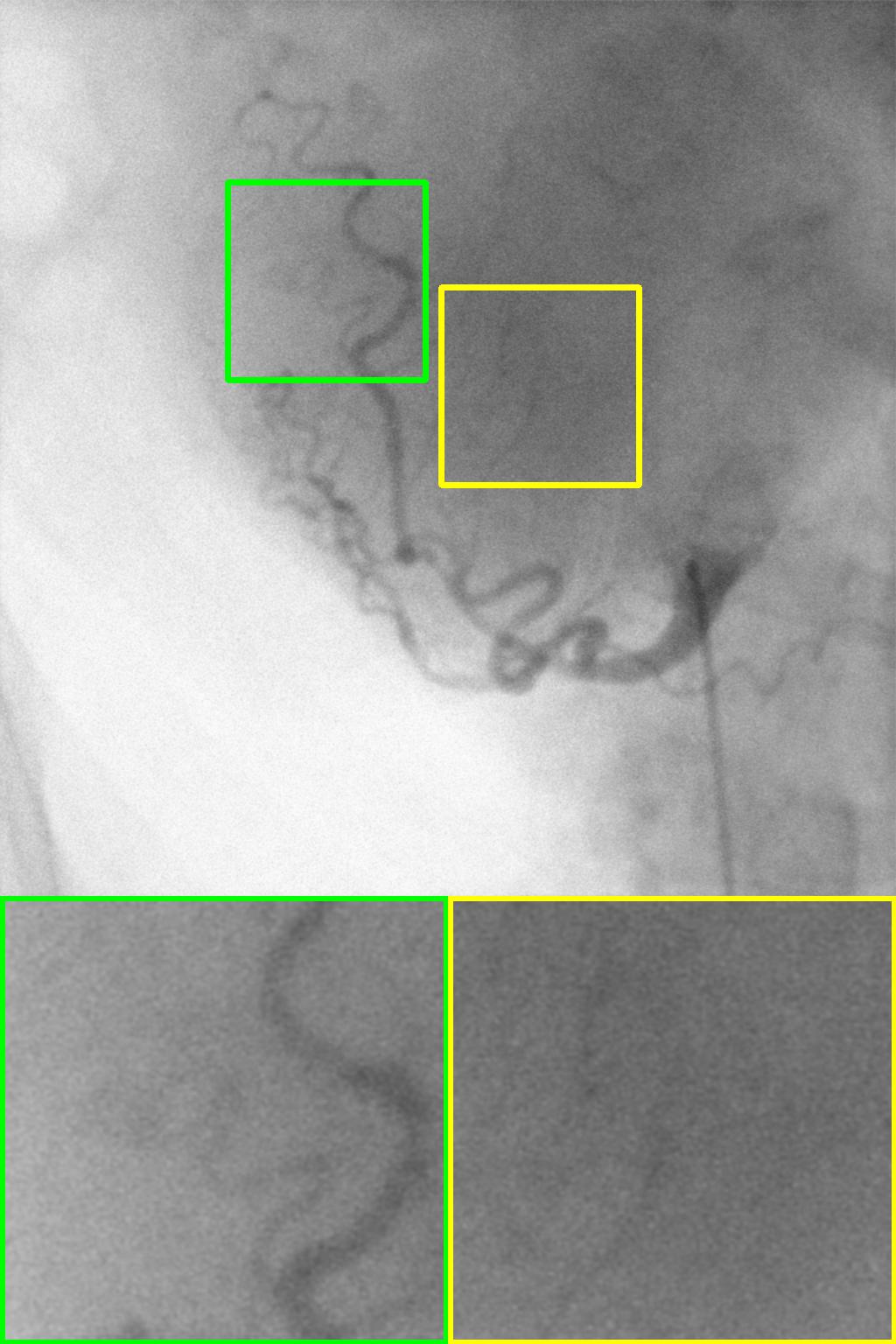}&
    \includegraphics[width=0.083\textwidth]{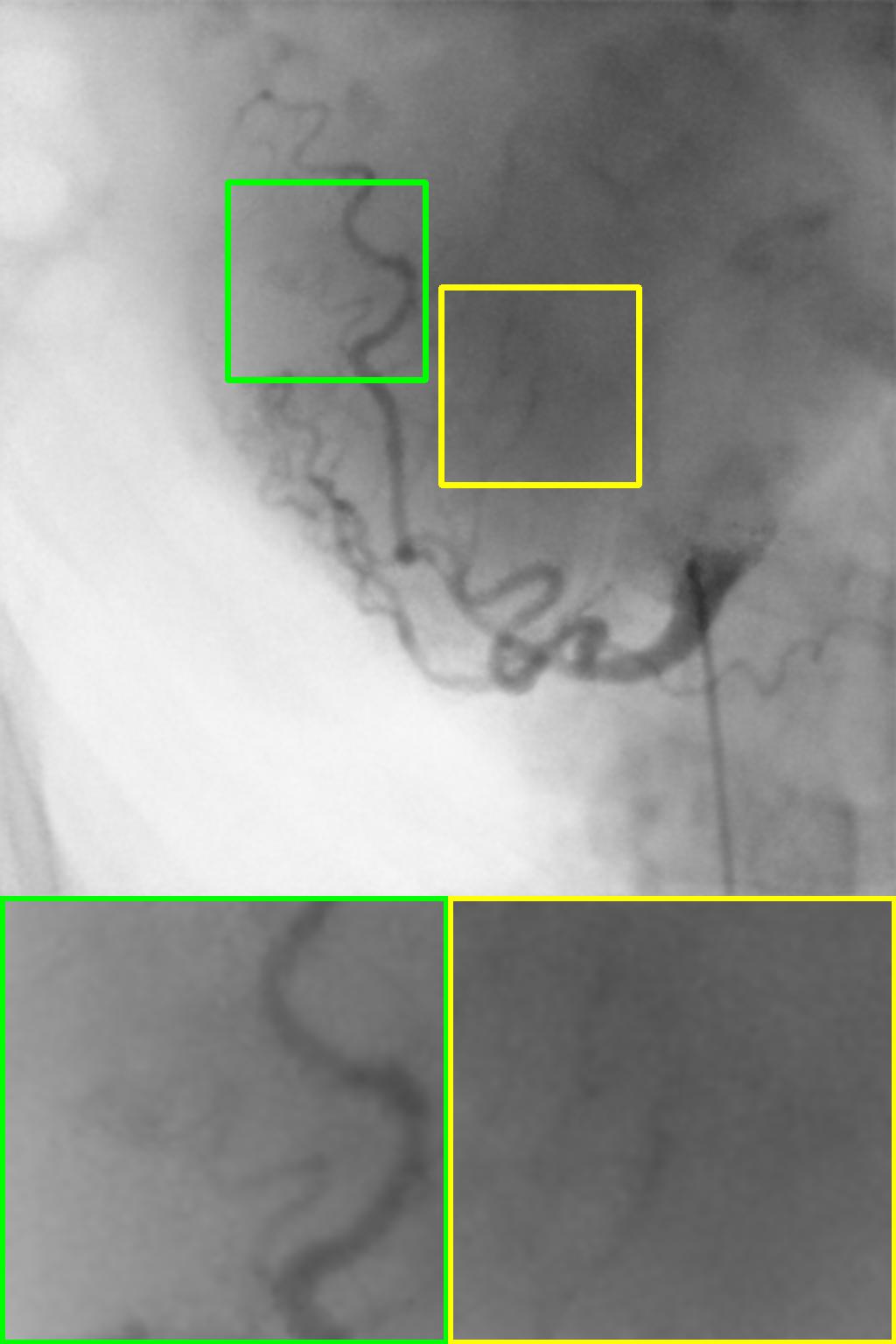}& 
    \\
    
\includegraphics[width=0.083\textwidth]{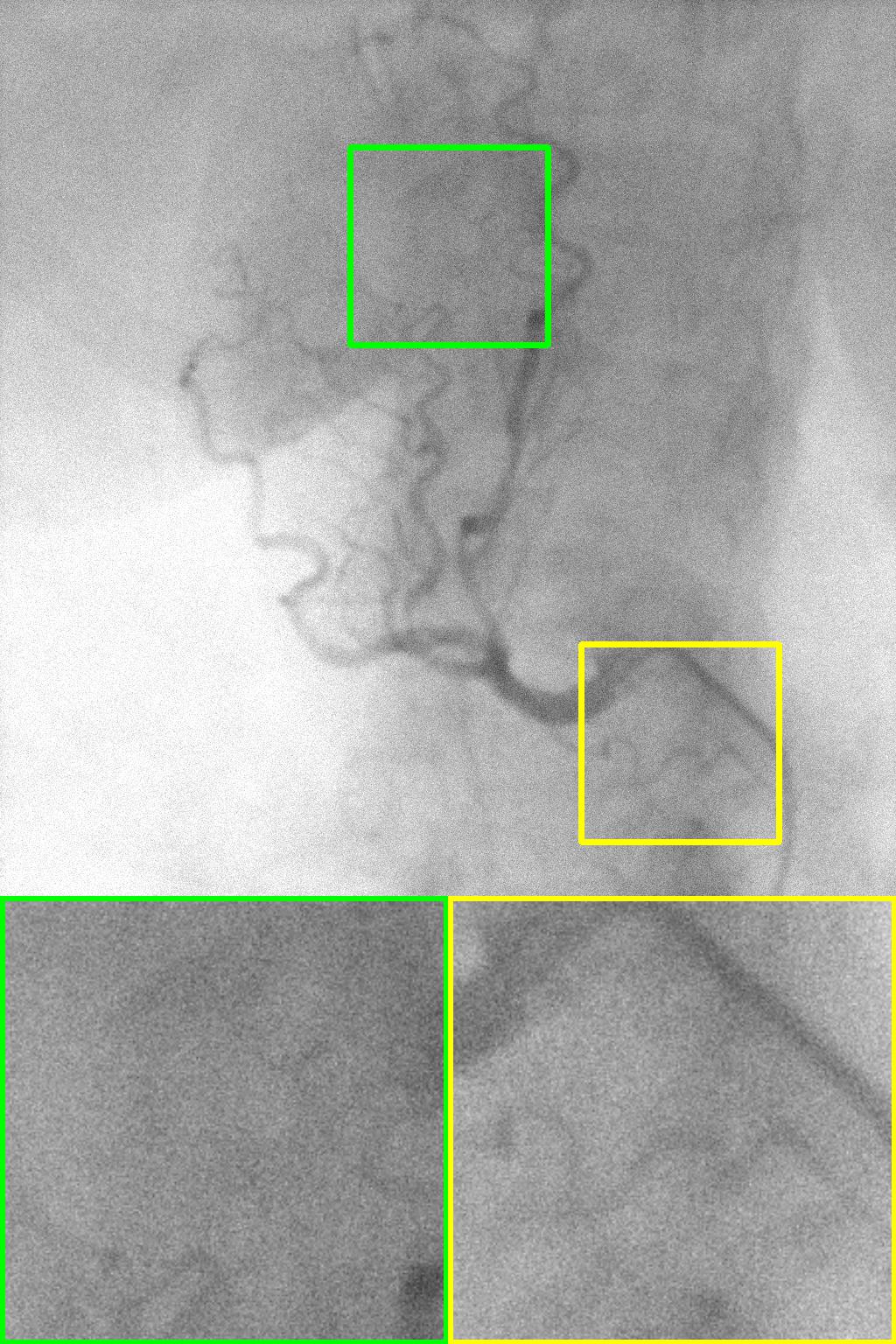} &
    \includegraphics[width=0.083\textwidth]{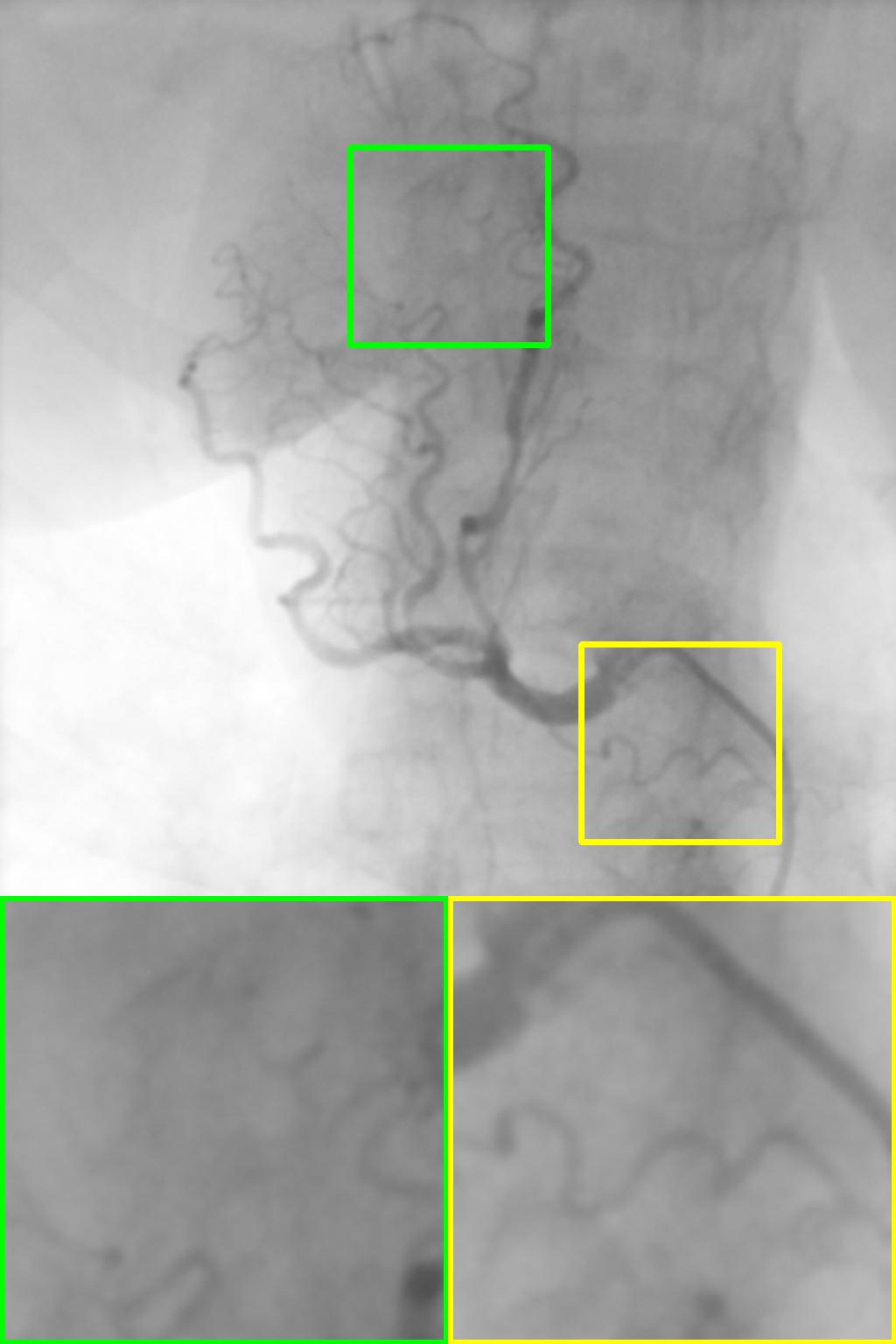} &
    \includegraphics[width=0.083\textwidth]{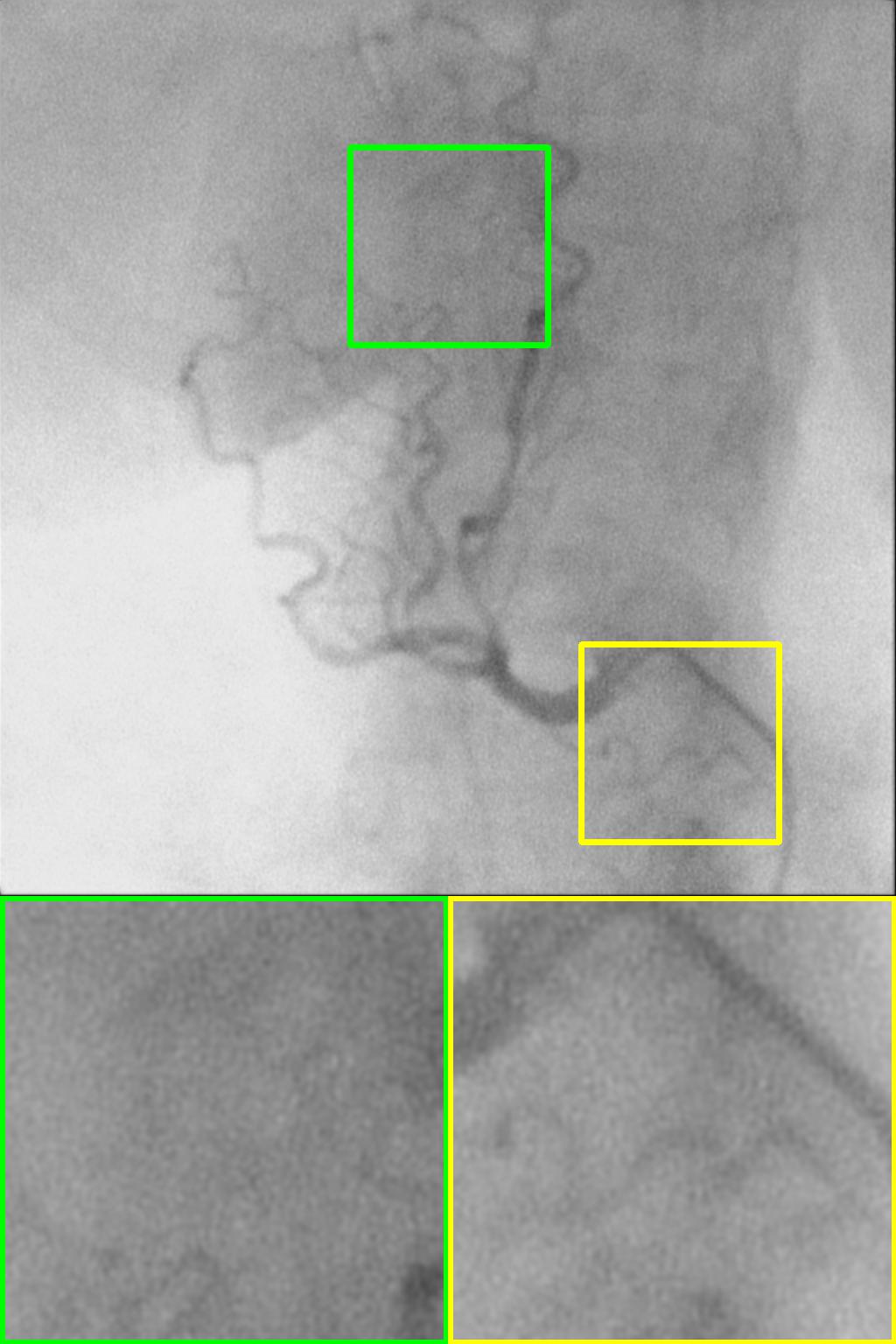}&
    \includegraphics[width=0.083\textwidth]{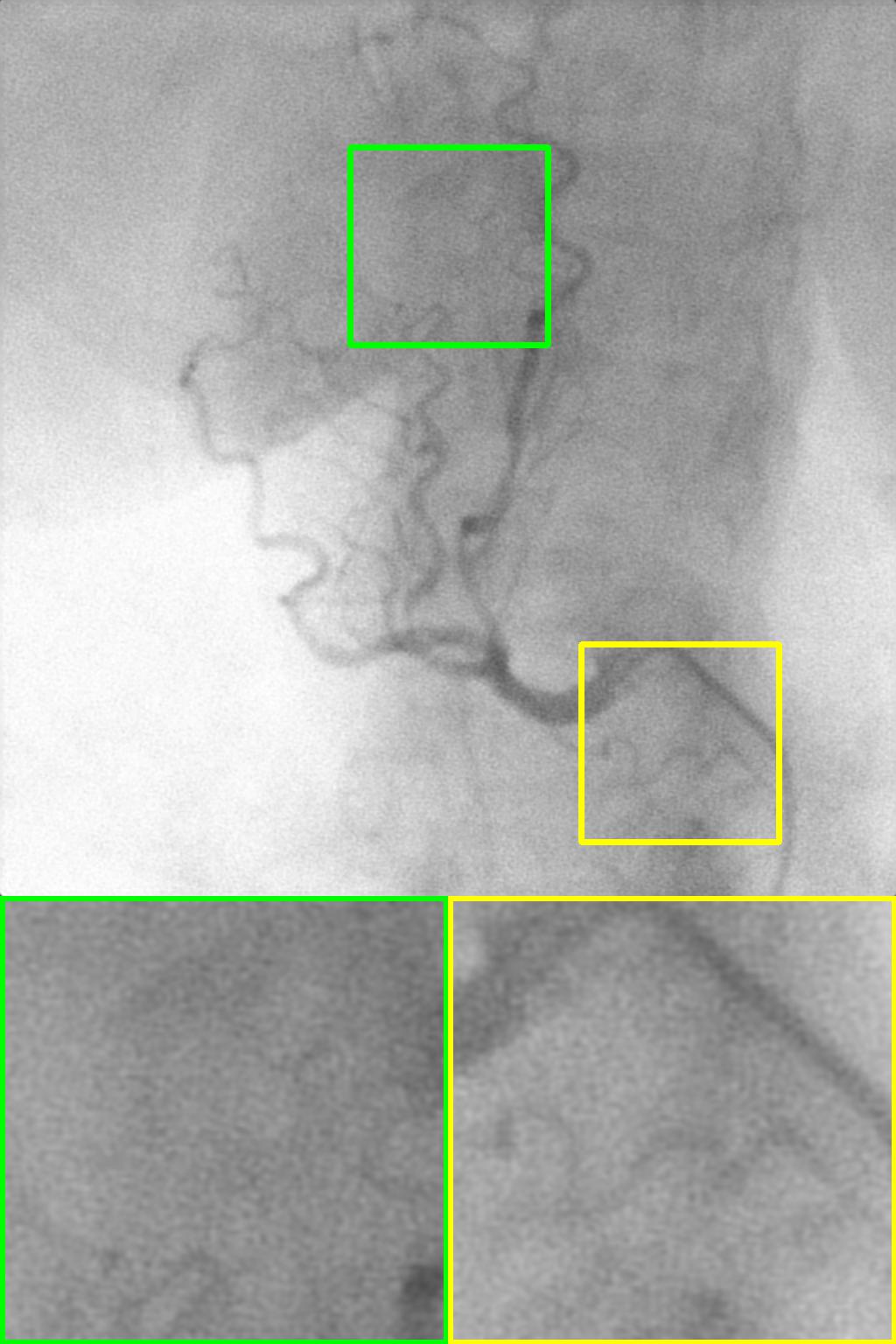}&
    \includegraphics[width=0.083\textwidth]{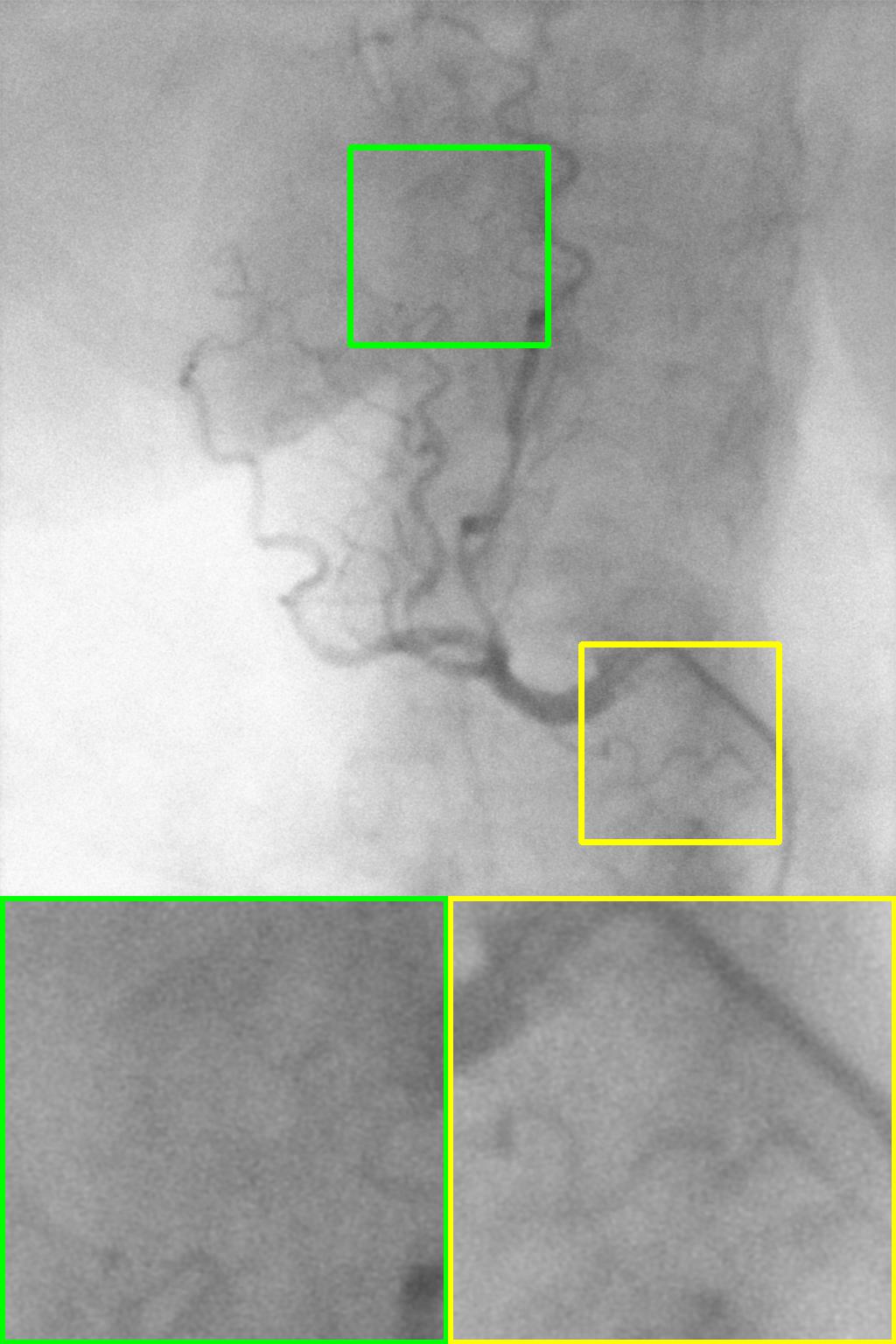}&
    \includegraphics[width=0.083\textwidth]{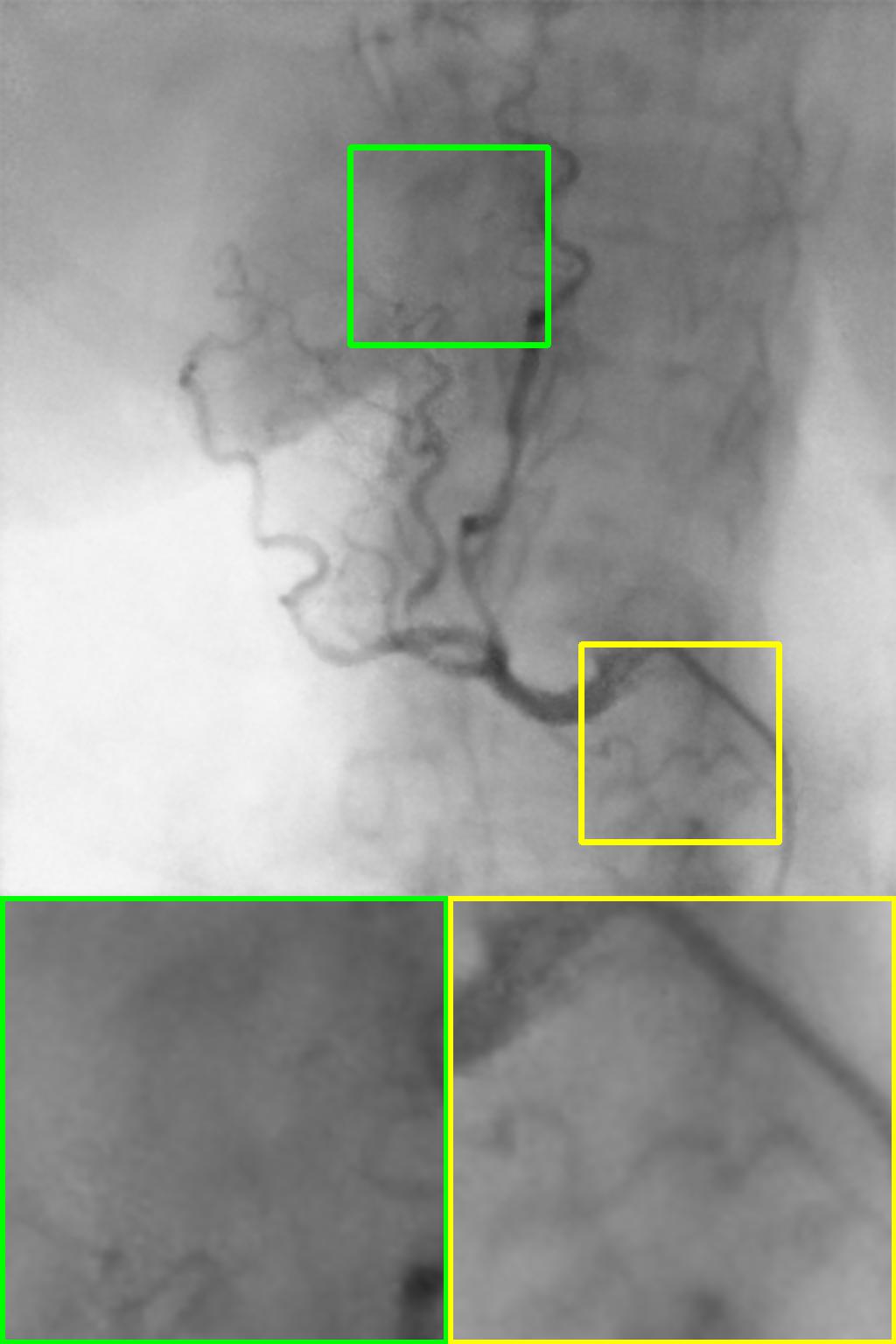}&
    \end{tabular}
}
\caption{Left: Visual comparison of denoising results. The images are with Gaussian noise ($\sigma= 0.003$). }
\label{fig:show-results} \vspace{-0.4cm}
\end{figure}

%% file: tables/clinical_show.tex
\begin{minipage}[t]{0.39\textwidth}
    \centering
    \small
    \includegraphics[width=26mm, height=17mm]{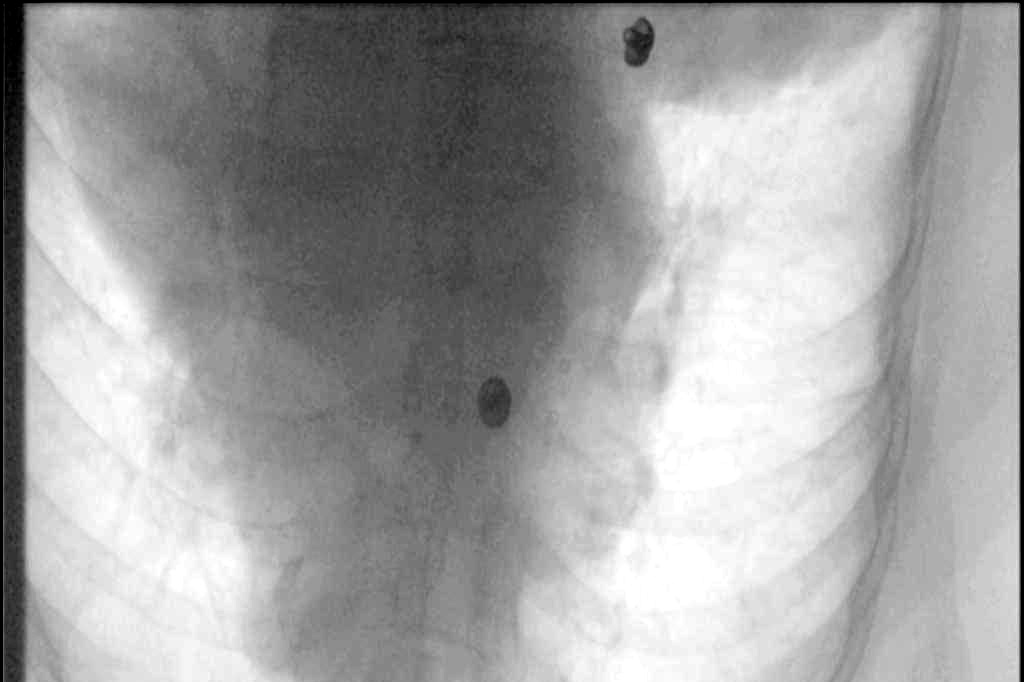}
    \includegraphics[width=26mm, height=17mm]{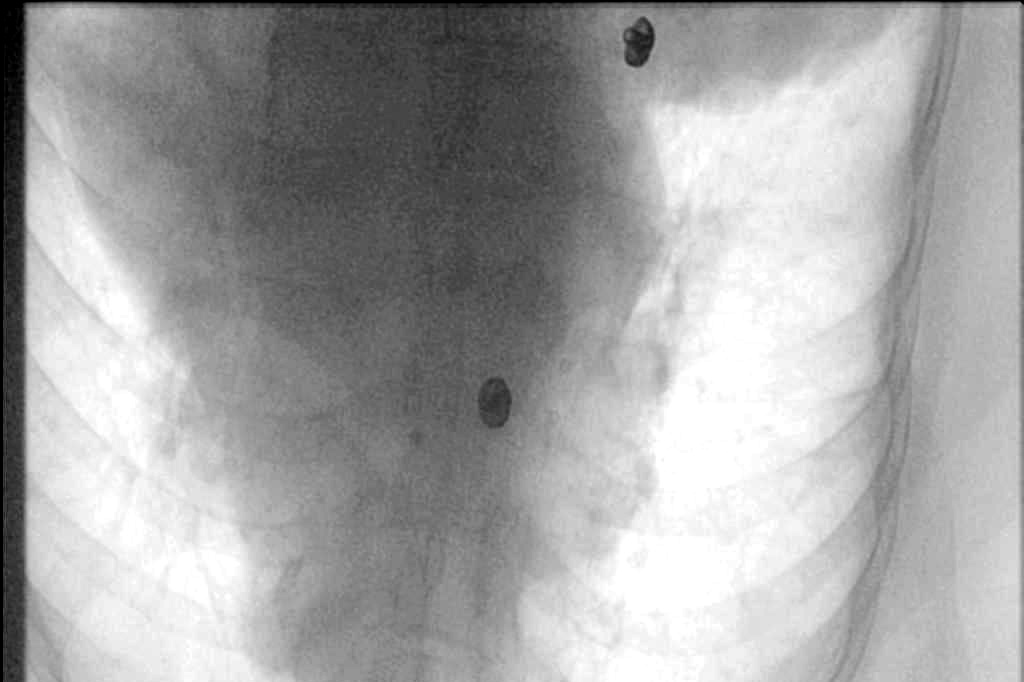}
\end{minipage}
\begin{minipage}[t]{0.49\textwidth}
    \centering
    \includegraphics[width=26mm, height=17mm]{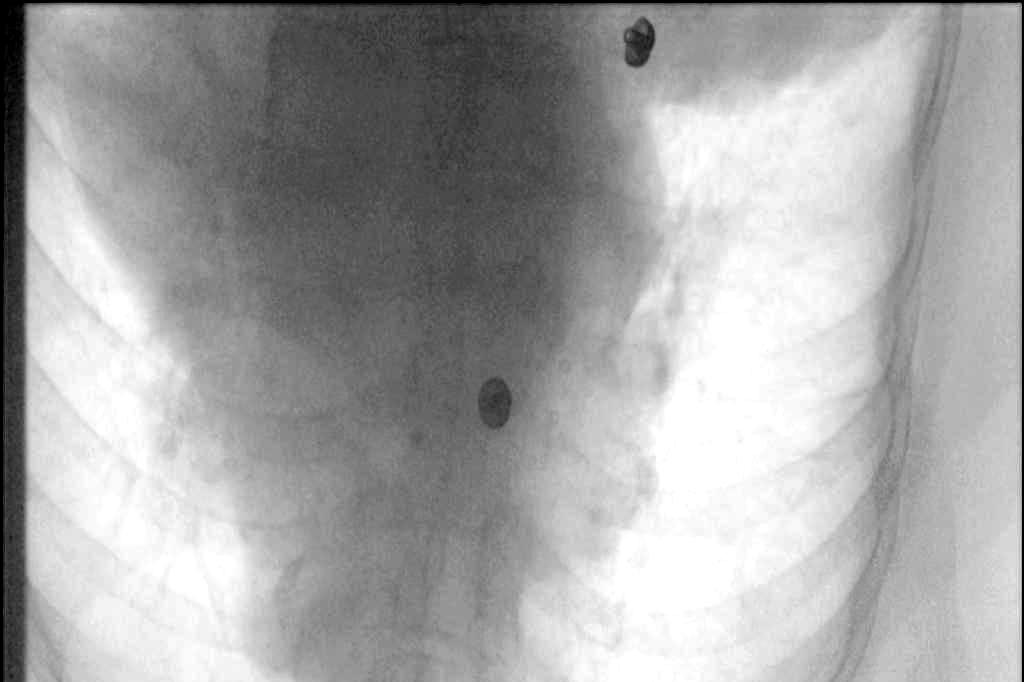}
    \includegraphics[width=26mm, height=17mm]{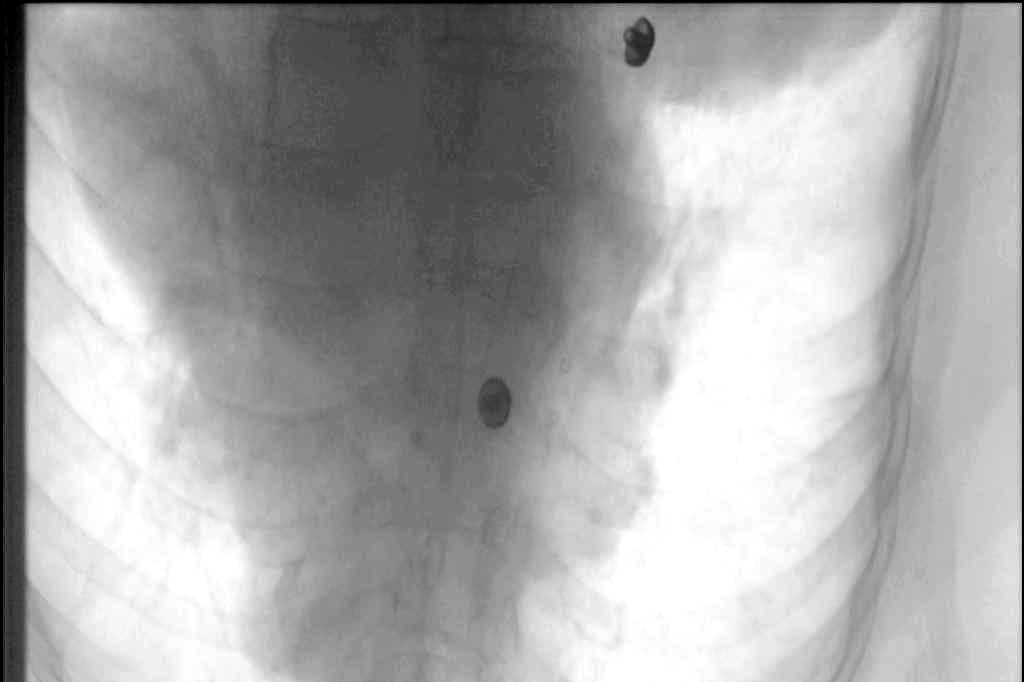}
\end{minipage}
\caption{Visualization of Clinical dataset.}
\label{fig: clinical}
\vspace{-0.5cm}

%% file: tables/clinical-table.tex




\tabcaption{Ratings results on Clinical dataset}
\centering
{
\begin{tabular}{|l|rr|}
\hline

\multicolumn{1}{|c|}{\multirow{2}{*}{}} & \multicolumn{2}{c|}{Clinical Dataset}   \\ \cline{2-3} 
\multicolumn{1}{|c|}{}                  & Ranking              & P-Value          \\ \hline
Raw                                     & 1.600±0.693          & \textless{}0.001 \\
Noise2Void                              & 3.300±0.962          & \textless{}0.001 \\
Noise2Self                              & 1.867±0.769          & \textless{}0.001 \\
Self2Self                               & 3.350±0.840          & \textless{}0.001 \\ \hline
Ours                                    & \textbf{4.883±0.584} & \textbf{nan}     \\ \hline
\end{tabular}

\label{tab:ranking}
}
\vspace{0.0cm}